\documentclass[aps, prd, reprint, superscriptaddress, nofootinbib,10pt]{revtex4-1}
\usepackage{graphicx}
\usepackage{xcolor}
\usepackage{bm}
\usepackage{mathtools}
\usepackage[colorlinks=true, filecolor=red, urlcolor=black, linkcolor=black, citecolor=red]{hyperref}
\usepackage{cleveref}
\usepackage{epigraph}
\begin{document}
\renewcommand{\abstractname}{\vspace{\baselineskip}} 
\title{The non-integrability of $L^{a,b,c}$ quiver gauge theories}
\author{Konstantinos S.~Rigatos}
\email{k.c.rigatos@soton.ac.uk}
\affiliation{
School  of  Physics  \&  Astronomy  and  STAG  Research  Centre,  University  of  Southampton, \\
Highfield, Southampton  SO$17$ $1$BJ, UK.\\
}
\begin{abstract}
We show that the $AdS_5 \times L^{a,b,c}$ solution in type IIB theory is non-integrable. To do so, we consider a string embedding and study its fluctuations which do not admit Liouville integrable solutions. We, also, perform a numerical analysis to study the time evolution of the string and compute the largest Lyapunov exponent. This analysis indicates that the string motion is chaotic. Finally, we consider the point-like limit of the string that corresponds to BPS mesons of the quiver theory. 
\end{abstract}
\setcounter{page}{1}\setcounter{footnote}{0}
\maketitle
\epigraph{This work is dedicated to the memory of David Graeber. Academia is much poorer without him.}{}
\section{P\lowercase{rolegomena}}
The gauge/string correspondence has evolved from the archetypical duality proposal \cite{Maldacena:1997re, Witten:1998qj, Gubser:1998bc} suggesting the equivalence of string theory in $AdS_5 \times S^5$ and the four dimensional $\mathcal{N}=4$ super Yang-Mills theory to more elaborate constructions with reduced amount of symmetry in an effort to probe toy models for field theories that appear in nature and gain intuition for the latter. One of the developments, to that end, involves the replacement of the five-dimensional sphere of the original $AdS_5 \times S^5$ background geometry by a five-dimensional Sasaki-Einstein manifold \footnote{For an excellent exposition and review on Sasaki-Einstein manifolds see \cite{Sparks:2010sn}.}, which we generically denote by $\mathcal{M}^{5}$ and therefore we obtain a duality between type IIB string theory on the AdS$_5 \times \mathcal{M}^{5}$ background and a quiver gauge theory that lives on the boundary \cite{Gubser:1998vd}. 

If we choose to specify the five-dimensional internal manifold to be $\mathcal{M}^{5} = T^{1,1}$ we obtain the so-called Klebanov-Witten model \cite{Klebanov:1998hh} which was the first one to be studied. However, nowadays, we have at our disposal more general (infinite) classes of such five (and also higher) dimensional manifolds which are characterized by either two or three indices and are denoted by $Y^{p,q}$ \cite{Gauntlett:2004yd} and $L^{a,b,c}$ \cite{Cvetic:2005ft, Martelli:2005wy}. These spaces possess a base topology that is $S^2 \times S^3$ and we know them explicitly in terms of metric descriptions. 

The boundary (dual) field theory descriptions have been obtained for both of the two different families of Sasaki-Einstein manifolds mentioned above. For the $Y^{p,q}$ manifolds the dual field theory description has been obtained \cite{Martelli:2004wu}. The holographic dual gauge theory description has also been obtained for the case of the $L^{a,b,c}$ manifolds \cite{Benvenuti:2005ja, Butti:2005sw,Franco:2005sm}. In this work, we will be concerned with the case of  the $L^{a,b,c}$ models. They are more general constructions and in fact it has been shown that the $Y^{p,q}$ manifolds can be obtained as special cases.

In a complementary approach towards the deeper understanding of gauge theories, an important role is played by integrability as its existence uncovers an affluent structure of conserved quantities. This in turn implies the solvability of the theory for any value of the gauge coupling. It is related to the previous discussion, since holographically we can associate the superstring worldsheet description to a field theory on the boundary without gravity, and therefore the integrability of the string side naturally becomes an equivalent statement for the integrability of the boundary gauge theory.

Integrability is present in the duality between the IIB theory in $AdS_5 \times S^5$ and the $\mathcal{N}=4$ SYM in the planar limit \cite{Beisert:2010jr}. It is only natural to ponder upon the possibility of whether or not we can discover new integrable structures in gauge theories with less symmetries. The classical integrability of the $AdS_5 \times S^5$ string is manifest, since the Lagrangian equations of motion can be expressed as a flat condition on the Lax connection \cite{Bena:2003wd}. Similar work to the above is also available for propagating strings in the Lunin-Maldacena background \cite{Lunin:2005jy}. This background is dual to the marginal Leigh-Strassler deformation, with a real parameter $\beta$, that preserves $\mathcal{N}=1$ supersymmetry, as was shown in \cite{Frolov:2005dj}. In the more general case where the $\beta$-deformation is complex, integrability is absent \cite{Frolov:2005ty, Berenstein:2004ys, Giataganas:2013dha}. 

While integrable field theories possess a number of appealing features, it is quite cumbersome to declare a certain theory integrable. This is due to the lack of a systematic approach in order to determine the Lax connection. Due to the aforementioned limitation, proving that a specific theory is non-integrable appears to be, in principle, a more wieldy problem. The full-fledged analysis consists of studying the non-linear PDEs that arise from the string $\sigma$-model. In practice, a facile approach is to study certain wrapped string embeddings and then analyse the resulting equations of motion. Since integrability has to be manifested universally, a single counter-example suffices to declare the full theory non-integrable.

One approach that has been undertaken in order to derive appropriate conditions of non-integrability is the S-matrix factorization on the worldsheet \citep{Wulff:2017lxh, Wulff:2017vhv, Wulff:2017hzy, Wulff:2019tzh}. A different, in spirit, approach was originally developed in \cite{Zayas:2010fs} and is based on the choice of a wrapped string embedding and the study of the relevant bosonic string $\sigma$-model. There is recent work \cite{Giataganas:2019xdj} on the relation between these two non-integrability approaches.

The procedure of \cite{Zayas:2010fs}, has been used subsequently in a series of papers \cite{Basu:2011di, Basu:2011fw, Stepanchuk:2012xi, Giataganas:2013dha, Nunez:2018qcj, Nunez:2018ags, Filippas:2019ihy, Filippas:2019bht, Filippas:2019puw, Giataganas:2017guj, Chervonyi:2013eja, Roychowdhury:2017vdo, Giataganas:2014hma, Roychowdhury:2019olt,Banerjee:2018ifm} that studied the classical (non)-integrability of different field theories. In a nutshell, the method consists of the following steps: write a string soliton that has $\mathcal{D}$ degrees of freedom and derive its equations of motion. Then find simple solutions for the $\left( \mathcal{D}-1\right)$ equations of motion. Replace in the final equation of motion these solutions and consider fluctuations. Thus we have arrived at a second-order linear differential equation which is called the normal variation equation (NVE) and is of the form $f'' + \mathcal{P} f' + \mathcal{Q} f = 0$. The existence or not of Liouville solutions is dictated by the mathematical approach developed by Kovacic \cite{KOVACIC19863}. If the result of the Kovacic method yields no Liouville integrable solutions or no solutions for the NVE, then we can declare the full theory as being a non-integrable. 

At this point we would like to stress that even if a background is characterised as being non-integrable in all generality, this does not preclude the existence of integrable subsectors in the theory. A very nice illustrative example of this situation is provided by the complex $\beta$-deformation. We have already mentioned that the complex $\beta$-deformation has been shown to be non-integrable in general, however the sub-sector that is comprised out of two holomorphic and one antiholomorphic scalar is known to be one-loop integrable \cite{Mansson:2007sh} as well as fast spinning strings in that subsector with a purely imaginary deformation parameter \cite{Puletti:2011hx}. Searching for integrable structures within non-integrable theories is an important question. It might provide useful links and further intuition for the transtition from integrable to non-integrable theories.

It is worthwhile mentioning that string solutions in the $L^{a,b,c}$ Sasaki-Einstein manifolds have been studied in \cite{Giataganas:2009dr} with a special emphasis on BPS configurations and different supersymmetric D-brane embeddings in the models have been studied in \cite{Canoura:2006es}. 

The structure of this work is as follows: we begin by briefly reviewing some basic facts regarding the $L^{a,b,c}$ metrics and subsequently we consider a string configuration positioned at the centre of the $AdS_5$ space and wrapping two angles of the $L^{a,b,c}$. We argue about the non-integrability of the field theory by studying the string dynamics. We find simple solutions of the equations of motion and allow the string to fluctuate around them. The study of the NVE does not yield a solution and thus we declare the quiver gauge theory to be generally non-integrable. We also perform a numerical analysis of the equations of motion governing the string embedding and compute the largest Lyapunov exponent. These numerical studies reveal chaotic dynamics of the string motion. We finally consider the point-like limit of the strings such that they are related to the BPS meson states of the field theory. 

\begin{figure}[h]
\centering
\includegraphics[scale=0.45]{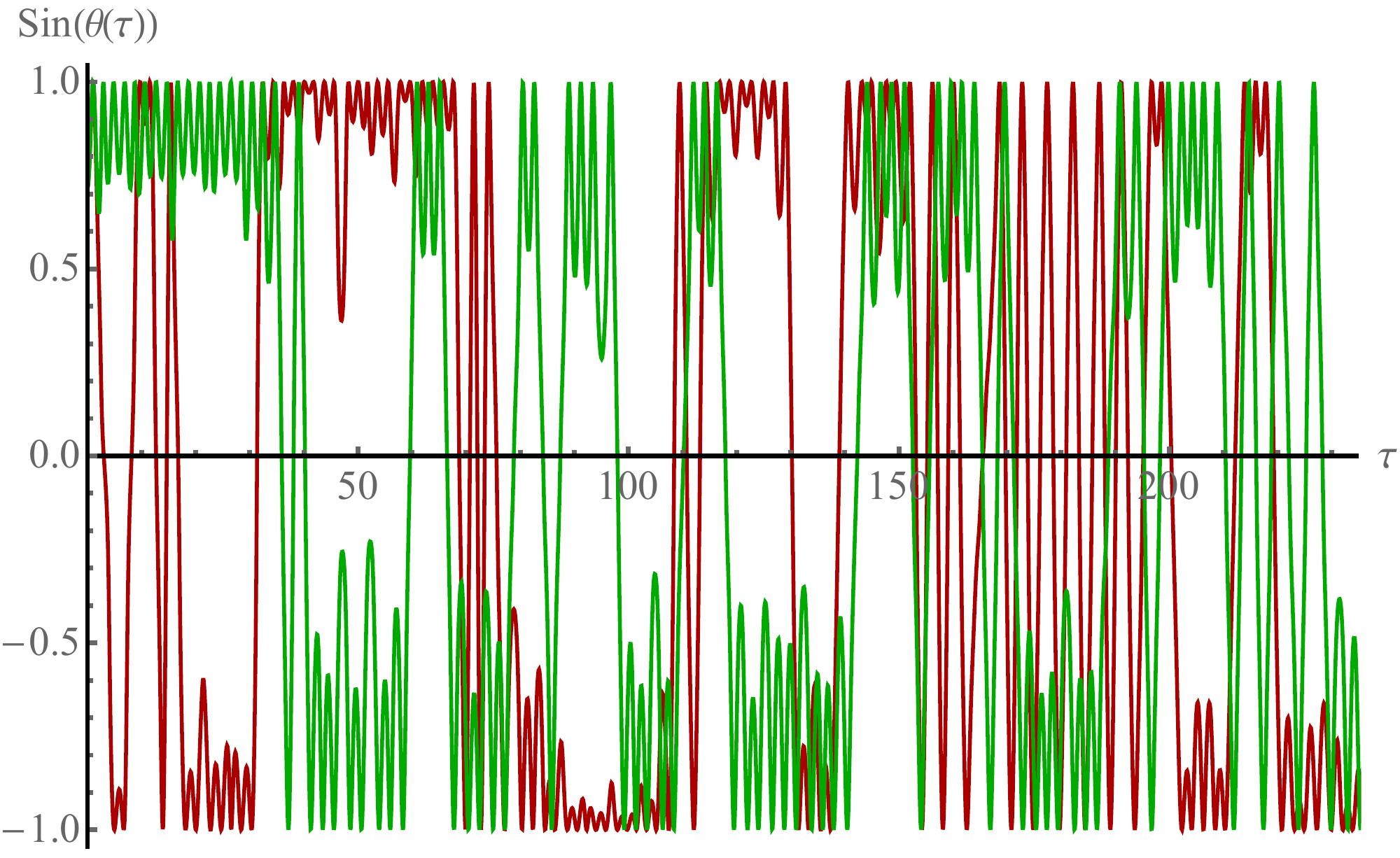} 
\centering
\caption{\label{fig: eleomsltn} The solution to the \cref{eq:thetaeom_1,eq:xeom_1} for the $L^{1,7,5}$ (red plot) and $L^{1,5,5}$ (green plot). The specific values for the coefficients $\alpha, \beta$ and the roots of $\Delta_x=0$ are discussed in the relevant section. The winding of the string in both cases is $\alpha_1=1, \alpha_2=2$. We have also set $\mu=1$ in both models. The time evolution indicates chaotic string motion.}
\end{figure}
\section{T\lowercase{he geometry}} \label{sec: geometry}
In this section we discuss the structure of $L^{a,b,c}$ spaces and for the reader's convenience, we quote the necessary relations to obtain the $Y^{p,q}$ manifolds from the $L^{a,b,c}$ geometries. 
	\subsection{The $L^{a,b,c}$ geometry}
The five-dimensional $L^{a,b,c}$ is written as 
\begin{equation}
ds^2_{L^{a,b,c}} = \left( d \zeta + \sigma \right)^2 +  ds^2_4,
\end{equation}
with the four-dimensional K\"ahler-Einstein metric
\begin{equation} \label{eq: Labc_part1}
\begin{aligned}
&ds^2_4 = \frac{\rho^2 dx^2}{4 \Delta_{x}}  + \frac{\rho^2 d \theta^2}{\Delta_{\theta}} + \frac{\Delta_{x}}{\rho^2} \left( \frac{\sin^2 \theta}{\alpha} d \phi + \frac{\cos^2 \theta}{\beta} d \psi  \right)^2 + \\
&\frac{\Delta_{\theta} \sin^2 \theta \cos^2 \theta}{\rho^2} \left[ \left(\frac{\alpha-x}{\alpha} \right) d \phi - \left(\frac{\beta-x}{\beta} \right) d \psi  \right]^2, 
\end{aligned}
\end{equation}
with the relevant quantities appearing above being given by
\begin{equation} \label{eq: Labc_part2}
\begin{split}
\sigma &= \left(\frac{\alpha-x}{\alpha} \right) \sin^2 \theta d \phi + \left(\frac{\beta-x}{\beta} \right) \cos^2 \theta d \psi, \\
\rho^2 &= \Delta_{\theta} - x, \\
\Delta_{x} &= x(\alpha-x)(\beta-x) - \mu, \\ 
\Delta_{\theta} &= \alpha \cos^2 \theta + \beta \sin^2 \theta.
\end{split}
\end{equation}
The metrics depend on two non-trivial parameters as anyone of the $\alpha, \beta, \mu$ can be set to any non-zero value by a rescaling of the other two. The toric principal orbits, $U(1) \times U(1) \times U(1)$, are degenerate when evaluated on the roots of $\Delta_x =0$ as well as at $\theta=0, \pi/2$. The ranges for the different coordinates are $0 \leq \theta \leq \pi/2$, $0 \leq \{\theta, \psi\} \leq 2\pi$ and the $x$-coordinate ranges from $x_1 \leq x \leq x_2$ with $x_{1,2}$ the smallest roots of the equation $\Delta_x =0$. The coordinate $\zeta$ is periodic and ranges $0\leq \zeta \leq \breve{\zeta}$ and $\breve{\zeta}$ is to be defined below. The three roots of the $\Delta_x =0$ equation are related to the constants $\alpha, \beta$ and $\mu$ of the metric in the following way
\begin{equation} \label{eq: manifolddef}
\begin{split}
\mu &= x_1 x_2 x_3, \quad \alpha + \beta = x_1 + x_2 + x_3, \\ 
\alpha \beta &= x_1 x_2 + x_1 x_3 + x_2 x_3
\end{split}
\end{equation}
where in the above $x_3$ is the third root of the aforementioned equation.

We can find relations for $x_1, x_2, \alpha, \beta$ in terms of the quantities $a, b, c, d$. They have been obtained in \cite{Canoura:2006es}, however we find it convenient and useful to repeat the analysis here . The normalized Killing vector fields are given by: 
\begin{equation}
\partial_{\phi}, \quad \partial_{\psi}, \quad \ell_{i} = A_i \partial_{\phi} + B_i \partial_{\psi} + C_i \partial_{\zeta} 
\end{equation}
with $i$ being valued either $1$ or $2$ and also, 
\begin{equation} \label{eq: defABC}
\begin{aligned}
A_i &= \frac{\alpha C_i}{x_i - \alpha} \\
B_i &= \frac{\beta C_i}{x_i - \beta} \\
C_i &= \frac{(\alpha - x_i)(\beta-x_i)}{2(\alpha+\beta)x_i-\alpha \beta - 3 x^2_i}
\end{aligned}
\end{equation} 

Now, we are at a position to give the value $\breve{\zeta}$ which is equal to
\begin{equation}
\breve{\zeta} =2\pi \frac{k |C_1|}{b}, \quad k=\gcd(a,b),
\end{equation}
and $d$ is defined to be:
\begin{equation} \label{eq: defd}
d = a+b-c
\end{equation}
The constants $A_i, B_i, C_i$ are related to to the integers $a,b,c$ that characterize the $L^{a,b,c}$ geometry through the relations
\begin{equation} \label{eq: preliminaries1} 
\begin{aligned}
a A_1 + b A_2 + c &= 0 \\
a B_1 + b B_2 +d &=0 \\
a C_1 + b C_2 &=
\end{aligned}
\end{equation}
A consequence of \cref{eq: preliminaries1} is that the ratios of $A_1 C_2 - A_2 C_1$, $B_1 C_2-B_2 C_1$, $C_1$, and $C_2$ have to be rational. More specifically, it has been shown that 
\begin{equation} \label{eq: defratios}
\begin{aligned}
\frac{c}{b} &= \frac{A_1 C_2 - A_2 C_1}{C_1} \\
\frac{d}{b} &= \frac{B_1 C_2 - B_2 C_1}{C_1} \\
\frac{b}{a} &= - \frac{C_1}{C_2}
\end{aligned}
\end{equation}
Using \cref{eq: manifolddef,eq: preliminaries1,eq: defratios} we can derive 
\begin{equation} \label{eq: usefuleqn}
\begin{aligned}
\frac{c}{b} &= \frac{x_1(x_3-x_1)}{x_2(x_3-x_2)} \\
\frac{a}{c} &= \frac{(\alpha - x_2)(x_3 - x_1)}{\alpha(\beta-x_1)} \\
\frac{c}{d} &= \frac{\alpha (\beta-x_1)(\beta-x_2)}{\beta (\alpha - x_1)(\alpha - x_2)} \\
\frac{c}{d} &=\frac{\alpha(x_3-\alpha)}{\beta(x_3-\beta)}
\end{aligned}
\end{equation}

We will keep the parameters $\alpha, \beta, \mu$ general and unspecified for most part of this work. However, in order to perform the numerical analysis of some equations we will need to specify them. In order to do that consistently, for a particular choice of $a,b,c$ that specifies the Sasaki-Einstein geometry, we determine $d$ using \cref{eq: defd}. We will set $\mu=1$ and use the first equation in \cref{eq: manifolddef} as well as the relations described in \cref{eq: usefuleqn} $\alpha, \beta, x_1, x_2, x_3$. We will give a specific example below when we analyze the Lagrangian equations of motion for an extended string. 
	\subsection{From $L^{a,b,c}$ to $Y^{p,q}$ spaces}
If we set $a+b=2c$, which in turn implies $\alpha=\beta$, the $L^{a,b,c}$ geometry reduces to the $Y^{p,q}$ spaces with the use of the following relations 
\begin{equation}
p-q=a, \quad p+q=b, \quad p=c.
\end{equation}
More explicitly, the transformation laws that reduce the $L^{a,b,c}$ metric to the $Y^{p,q}$ one are \cite{Butti:2005sw}
\begin{equation}
\begin{split}
\bar{\psi} &= 3 \zeta + \psi + \phi, \quad \bar{\phi} = \phi-\psi, \quad \bar{\beta} = -(\phi+\psi), \\
\bar{\theta} &= 2 \theta, \quad \bar{y} = \frac{3x-\alpha}{2 \alpha}.
\end{split}
\end{equation}
and the $Y^{p,q}$ metric is written explicitly as 
\begin{equation}
ds^2_{Y^{p,q}} =  \left( \frac{1}{3} d \bar{\psi} + \bar{\sigma} \right)^2 +  d\bar{s}^2_4,
\end{equation}
where in the above $ \bar{\sigma}$ is given by:
\begin{equation}
\bar{\sigma} = -\frac{1}{3} \left( \cos \bar{\theta} ~ d\bar{\phi} + \bar{y} d \bar{\beta} + \bar{c} \cos \bar{\theta} ~ d\bar{\phi} \right),
\end{equation}
with the four-dimensional metric of the $Y^{p,q}$ space is 
\begin{equation}
\begin{split}
d\bar{s}^2_4 = &\frac{1 - \bar{c} \bar{y}}{6} \left( d \bar{\theta}^2 + \sin^2 \bar{\theta} ~ d\bar{\phi}^2 \right) + \frac{1}{w(\bar{y}) q(\bar{y})} d\bar{y}^2 + \\ &\frac{w(\bar{y}) q(\bar{y})}{36} \left( d\bar{\beta} + \bar{c} \cos \bar{\theta} ~ d\bar{\phi} \right)^2 ,
\end{split}
\end{equation}
and finally the functions $w$ and $q$ have the form
\begin{equation}
w(\bar{y}) = \frac{2(a-\bar{y}^2)}{1- \bar{c} \bar{y}}, \quad q(\bar{y}) = \frac{a-3\bar{y}^2 + 2 \bar{c} \bar{y}^3}{a-\bar{y}^2}.
\end{equation}

A comment is in order here. We saw that for special values of the $\{a,b,c\}$ parameters the $L^{a,b,c}$ models reduce to the $Y^{p,q}$ ones, which are known to be non-integrable \cite{Basu:2011fw}. However, the $\{p,q \}$ parameter space is much smaller and only a subset of all the possible choices of the full space spanned by $\{a,b,c\}$, corresponding to the theories we are examining here.
	\subsection{The $AdS_5 \times L^{a,b,c}$ geometry}
The geometry that we want to consider is given by 
\begin{equation} \label{eq: 10_dim_geometry}
ds^2_{10} =  L^2  ds^2_{AdS_5} + L^2 ~ ds^2_{L^{a,b,c}}. 
\end{equation}
For our purposes it is most convenient to describe the five-dimensional $AdS$ space using global coordinates, 
\begin{equation}
ds^2_{AdS_5} =  -\cosh^2 \varrho ~ dt^2 + d \varrho^2 + \sinh^2 \varrho ~ d\Omega^2_3.
\end{equation}
In the above, $d\Omega^2_3$ is the round metric of a unit three-sphere which is given explicitly by  
\begin{equation}
d\Omega^2_3 = dw^2_1 + \sin^2 w_1 dw^2_2 + \sin^2 w_1  \sin^2 w_2 dw^2_3.
\end{equation}
The angles are valued within the ranges $0 \leq w_1 \leq \pi/2$ and $0 \leq w_2,w_3 \leq 2 \pi$.
\section{S\lowercase{tring dynamics}} \label{sec: dynamics}
The Polyakov action is given by
\begin{equation} \label{eq: polyakov_action}
S = - \frac{1}{4 \pi \alpha'} \int d^2 \sigma ~ h^{\alpha \beta} ~ G_{MN} ~ \partial_{\alpha} X^{M} \partial_{\beta} X^{N}
\end{equation}
in the conformal gauge and must be supplemented by the Virasoro constraints 
\begin{equation} \label{eq: virasoro_constraints}
\begin{split}
T_{\tau \sigma} = T_{\sigma \tau} =&G_{MN} \dot{X}^{M} \acute{X}^{N} = 0, \\
2T_{\tau \tau} = 2T_{\sigma \sigma} =&G_{MN} \left( \dot{X}^{M} \dot{X}^{N} + \acute{X}^{M} \acute{X}^{N} \right) = 0
\end{split}
\end{equation}
where we have used the abbreviations $\dot{X} \equiv \partial_{\tau} X$ and $ \acute{X} \equiv  \partial_{\sigma} X$.

Since the string motion in the $AdS_5 \times S^5$ background is integrable and there are no NS-fluxes to deform the $\sigma$-model in our case of interest, non-integrability will be manifested in the structure of the $L^{a,b,c}$ manifold. Thus, a natural choice for the string embedding is to localize the classical string that we want to study at the centre of the $AdS_5$ space ($\varrho=0$) and then wrap two directions of the $L^{a,b,c}$ space, more specifically the $\phi$ and $\psi$. Explicitly, we are using the ansatz: 
\begin{equation} \label{eq: string_soliton}
\begin{aligned}
t&=t(\tau), 				& 	x&=x(\tau), 				& \phi&=\alpha_{1} \sigma \\
\zeta&=\zeta(\tau), 	& 	\theta&=\theta(\tau), 	& \psi&=\alpha_{2} \sigma.
\end{aligned}
\end{equation}

Note that the string configuration described in \cref{eq: string_soliton} is similar in spirit as the one used for the $T^{1,1}$ \cite{Basu:2011di} as well as the $T^{p,q}$ and $Y^{p,q}$ models \cite{Basu:2011fw}. 
	\subsection{W\lowercase{rapped strings at the centre of} $AdS_5$}
We can now evaluate the Lagrangian density of the $\sigma$-model for our particular choice of the string embedding described by \cref{eq: string_soliton},
\begin{equation} \label{eq: string_lagrangian}
\begin{aligned}
&\mathcal{L} = \dot{t}^2 - \dot{\zeta}^2 - \frac{\rho^2}{4} \left(  \frac{\dot{x}^2}{\Delta_x} + 4 \frac{\dot{\theta}^2}{\Delta_{\theta}} \right) + \mathcal{A}_1~\sin^2 \theta~\alpha^2_1 \\
&+ \mathcal{A}_2 \cos^2 \theta \alpha^2_2 +  \mathcal{A}_3 \sin^2(2\theta) \alpha_1 \alpha_2
\end{aligned}
\end{equation}
where in the above the prefactors $\mathcal{A}_{1,2,3}$ are given by
\begin{equation}
\begin{aligned}
\mathcal{A}_{1} = &\left(\frac{\alpha-x}{\alpha} \right)^2 \sin^2 \theta \\
&+ \frac{\Delta_x \sin^2 \theta +  \Delta_{\theta} \cos^2 \theta (\alpha-x)^2}{\alpha^2 \rho^2},									\\
\mathcal{A}_{2} = &\left(\frac{\beta-x}{\beta} \right)^2 \cos^2 \theta \\
&+ \frac{\Delta_x \cos^2 \theta + \Delta_{\theta} \sin^2 \theta (\beta-x)^2}{\beta^2 \rho^2}, 									\\
\mathcal{A}_{3} = &\frac{\Delta_x -(x-\alpha)(x-\beta)(\Delta_{\theta}-\rho^2)}{2 \alpha \beta \rho^2}\, .
\end{aligned}
\end{equation}
The equations of motion that follow from the Lagrangian read
\begin{subequations} 
\begin{equation}
\ddot{t} = 0, \label{eq:teom_1}
\end{equation}
\begin{equation}
\ddot{\zeta} = 0, \label{eq:zetaeom_1}
\end{equation}
\begin{equation}
\begin{aligned}
&8\left(1 - \frac{2 x}{\alpha+\beta+(\alpha-\beta) \cos(2\theta)} \right) \ddot{\theta} = \\
&(-\alpha + \beta) \sin(2 \theta) \left( \frac{\dot{x}^2}{\Delta_x} - \frac{4 x \dot{\theta}^2}{\Delta^2_{\theta}} \right) + \frac{8}{\Delta_{\theta}} \dot{\theta} \dot{x}  \\
&- 4 \frac{\sin(2 \theta)}{(-\alpha + \beta)} \left( \mathcal{B}_1 \alpha^2_1 + \mathcal{B}_2 \alpha^2_2 - \frac{\mu}{\alpha \beta} \mathcal{B}_3 \alpha_1 \alpha_2 \right), \label{eq:thetaeom_1}
\end{aligned}
\end{equation}
\begin{equation}
\begin{aligned}
&2 \frac{\Delta_{\theta}-x}{\Delta_{x}} \ddot{x} = 2 \frac{\dot{x}+(\alpha-\beta)\sin(2\theta)\dot{\theta}}{\Delta_x} \dot{x} + \frac{4}{\Delta_{\theta}} \dot{\theta}^2   \\
&- \frac{1}{\Delta_x} \left(\vphantom{\frac{1}{2}} 1 - \frac{\Delta_{\theta}-x}{\Delta_x}  \right. \\
&\left. (\alpha \beta -2 (\alpha+\beta)x+3x^2) \vphantom{\frac{1}{2}} \right) \dot{x}^2  \\
&+ \frac{1}{2(\Delta_{\theta}-x)^2} \\
&\left(\vphantom{\frac{1}{2}} \mathcal{C}_1 \sin^2 \theta \alpha^2_1+ \mathcal{C}_2 \cos^2 \theta \alpha^2_2 - \frac{\mu}{\alpha \beta} \mathcal{C}_3 \alpha_1 \alpha_2 \right).  \label{eq:xeom_1}
\end{aligned}
\end{equation}
\end{subequations}
In the above equations the $\mathcal{B}$-prefactors are explicitly given by: 
\begin{subequations}
\begin{equation}
\begin{aligned}
&\mathcal{B}_1 = -\frac{\mu }{\alpha ^2}-\alpha +\beta -\frac{\beta  x}{\alpha }+x \\
&+ \frac{4 \mu  (\alpha -x)^2}{\alpha ^2 ((\alpha -\beta ) \cos (2 \theta )+\alpha +\beta -2 x)^2},
\end{aligned}
\end{equation}
\begin{equation}
\begin{aligned}
&\mathcal{B}_2 = \alpha -\frac{\mu }{\beta ^2}-\beta-\frac{\alpha  x}{\beta }+x \\
&+\frac{4 \mu  (\beta -x)^2}{\beta ^2 ((\alpha -\beta ) \cos (2 \theta )+\alpha +\beta -2 x)^2},
\end{aligned}
\end{equation}
\begin{equation}
\mathcal{B}_3 =-1 + \frac{4 (\alpha -x) (\beta -x)}{((\alpha -\beta ) \cos (2 \theta )+\alpha +\beta -2 x)^2},
\end{equation}
\end{subequations}
while the $\mathcal{C}$-prefactors are equal to 
\begin{subequations}
\begin{equation}
\begin{aligned}
&\mathcal{C}_1 = -\frac{4 \mu  \cos (2 \theta )}{\alpha ^2}+\frac{4 \mu }{\alpha ^2}-\frac{4 \beta ^2 \cos (2 \theta )}{\alpha }\\
&+\frac{\beta ^2 \cos (4 \theta )}{\alpha }+\frac{3 \beta^2}{\alpha }+4 \alpha  \cos (2 \theta )+\alpha  \cos (4 \theta ) \\
&+3 \alpha -2 \beta  \cos (4 \theta )+2 \beta +\frac{8 x^2}{\alpha }+\frac{8 \beta  x \cos (2 \theta)}{\alpha } \\
&- \frac{8 \beta  x}{\alpha }-8 x \cos (2 \theta )-8 x, 
\end{aligned}
\end{equation}
\begin{equation}
\begin{aligned}
&\mathcal{C}_2=\frac{4 \alpha ^2 \cos (2 \theta )}{\beta }+\frac{\alpha ^2 \cos (4 \theta )}{\beta }+\frac{3 \alpha ^2}{\beta }\\
&-2 \alpha  \cos (4 \theta )+2 \alpha +\frac{4 \mu  \cos(2 \theta )}{\beta ^2}+\frac{4 \mu }{\beta ^2} \\ 
&-4 \beta  \cos (2 \theta ) + \beta  \cos (4 \theta ) +3 \beta +\frac{8 x^2}{\beta }\\
&-\frac{8 \alpha  x \cos (2 \theta)}{\beta }-\frac{8 \alpha  x}{\beta }+8 x \cos (2 \theta )-8 x,
\end{aligned}
\end{equation}
\begin{equation}
\mathcal{C}_3 = -2 \sin^2(2 \theta).
\end{equation}
\end{subequations}
From the above, the equations \cref{eq:teom_1,eq:zetaeom_1} can be integrated immediately 
\begin{equation}
\dot{t}^2 = E^2, \quad \dot{\zeta}^2 = J^2\, ,
\end{equation}
with $E$ and $J$ being constants.

The equations of motion above, \cref{eq:teom_1,eq:zetaeom_1,eq:thetaeom_1,eq:xeom_1}, are constrained by the Virasoro conditions. 
We evaluate \cref{eq: virasoro_constraints} for our particular string conifguration \cref{eq: string_soliton}
\begin{subequations} \label{eq: virasoro_spec}
\begin{equation} \label{eq: virasoro_spec_1}
\begin{aligned}
&2T_{\tau \tau} = 2T_{\sigma \sigma} = - \dot{t}^2 + \dot{\zeta}^2 + \frac{\rho^2}{4} \left(  \frac{\dot{x}^2}{\Delta_x} + 4 \frac{\dot{\theta}^2}{\Delta_{\theta}} \right)  \\
&+\mathcal{A}_1 \sin^2 \theta \alpha^2_1 +\mathcal{A}_2 \cos^2 \theta \alpha^2_2 \\
&+  \mathcal{A}_3 \sin^2(2\theta) \alpha_1 \alpha_2 = 0,
\end{aligned}
\end{equation}
\begin{equation} \label{eq: virasoro_spec_2}
\begin{aligned}
T_{\tau \sigma}=T_{\sigma \tau} = &\left( \frac{\alpha-x}{\alpha} \sin^2 \theta \alpha_1 + \right. \\
&\left. \frac{\beta-x}{\beta} \cos^2 \theta \alpha_2 \right) \dot{\zeta} = 0.
\end{aligned}
\end{equation}
\end{subequations}

We can express the theory under consideration in a Hamiltonian formalism. The conjugate momenta are given by 
\begin{equation}
\begin{aligned}
p_t &= 2 \dot{t}, \quad 									&p_{\zeta} &= -2 \dot{\zeta}, \\
p_x&=-\frac{\rho^2}{2 \Delta_x} \dot{x}, \quad 	&p_{\theta} &= - \frac{2 \rho^2}{\Delta_{\theta}} \dot{\theta},
\end{aligned}
\end{equation}
and the Hamiltonian density is equal to
\begin{equation}
\begin{aligned}
&\mathcal{H} = \frac{1}{4 \rho^2} \left[ \rho^2 (p^2_t - p^2_{\zeta}) - 4 \Delta_x p^2_x - \Delta_{\theta} p^2_{\theta} \vphantom{\frac{1}{2}} \right]  \\ 
&- \mathcal{A}_1~\sin^2 \theta~\alpha^2_1- \mathcal{A}_2 \cos^2 \theta \alpha^2_2 - \mathcal{A}_3 \sin^2(2\theta) \alpha_1 \alpha_2.
\end{aligned}
\end{equation}
The equations of motion that follow from the Hamiltonian are, of course, identical with the E\"uler-Lagrange equations \cref{eq:teom_1,eq:zetaeom_1,eq:thetaeom_1,eq:xeom_1}. 
		\subsubsection{\textbf{F\lowercase{luctuations around the simple solutions}}}
The $\theta$ and $x$ equations of motion are coupled \cref{eq:thetaeom_1,eq:xeom_1}. However, to prove the non-integrability of extended string motion we can simplify this situation by freezing one dimension and fluctuating the other around a simple solution. 

To that end, it is easy to see that there exists an obvious and simple solution to the equation of motion for $\theta(\tau)$ \cref{eq:thetaeom_1} which is given by
\begin{equation}
\theta = \dot{\theta} = \ddot{\theta} = 0\, .
\end{equation}
We refer to it as the straight line solution. Using the above, the equation of motion for $x(\tau)$, \cref{eq:xeom_1}, simplifies to 
\begin{equation}
\begin{aligned}
&2 \frac{x-\alpha}{\Delta_x}\ddot{x} = -\frac{4}{\beta} \left(1+\frac{\mu}{\beta(\alpha-x)^2} \right) + \frac{\dot{x}^2}{\Delta^2_x} \\
&(2x^3-(4\alpha+\beta)x^2+2\alpha(\alpha+\beta)x-\alpha^2 \beta+\mu)  = 0\, .
\end{aligned}
\end{equation}
Let us denote the solution to the above equation by $\bar{x}$, where we have omitted the explicit time dependence for notational convenience. 

We fluctuate now the $x$ coordinate around that particular solution as $x = \bar{x} + \varepsilon \mathcal{X}$ with $\varepsilon \rightarrow 0$ while keeping the $\theta$ coordinate frozen according to $\{ \theta = \dot{\theta} = \ddot{\theta} = 0 \}$. We work to linear order in the small parameter $\varepsilon$ and the resulting equation is the NVE for the $x$-coordinate. It reads:
\begin{equation} \label{eq: NVE_x}
\begin{aligned}
\ddot{\mathcal{X}} + &\mathcal{P} ~ \dot{\bar{x}} ~ \dot{\mathcal{X}} + \frac{1}{2} \left( \vphantom{\frac{1}{2}} (\beta-\bar{x})\bar{x} \right. \\
&\left.+ \frac{\mu}{\bar{x}-\alpha} \right) \left( \mathcal{Q}_1 \ddot{\bar{x}} + \mathcal{Q}_2 \dot{\bar{x}}^2 + \mathcal{Q}_3 \right) \mathcal{X} = 0
\end{aligned}
\end{equation}
with the prefactors being given by:
\begin{equation}
\begin{aligned}
&\mathcal{P} = \frac{\alpha^2 \beta - \mu + \bar{x} (-2\alpha(\alpha+\beta)+(4\alpha+\beta-2\bar{x}))\bar{x}}{(\alpha-\bar{x})(\mu + (\alpha-\bar{x})(-\beta+\bar{x})\bar{x})}  \\
&\mathcal{Q}_1 = \frac{2}{((\alpha-\bar{x})(\mu + (\alpha-\bar{x})(-\beta+\bar{x})\bar{x}))^2} (-\alpha^2 \beta \\
& + \mu + 2 \alpha (\alpha + \beta)\bar{x} - (4\alpha+\beta)\bar{x}^2+2\bar{x}^3)\\
&\mathcal{Q}_2 = \frac{2 ~ \mathcal{N}}{((\alpha-\bar{x})(\mu + (\alpha-\bar{x})(-\beta+\bar{x})\bar{x}))^3} \\
&\mathcal{N} = \left[\vphantom{\frac{1}{2}} -\alpha^3 \beta^2  + \alpha (\alpha + 2 \beta) \mu + \bar{x} ( 3 \alpha^2 \beta (\alpha + \beta)  \right. \\
&\left.- 3 (2 \alpha + \beta) \mu +  \bar{x} (-3\alpha(\alpha^2+3 \alpha \beta + \beta^2) + 6 \mu \right. \\
&\left.+\bar{x} ( 9 \alpha^2 + 9 \alpha \beta + \beta^2 + 3 \bar{x}  (-3 \alpha - \beta + \bar{x}))))
 \vphantom{\frac{1}{2}} \right] \\
&\mathcal{Q}_3 = - \frac{8 \mu}{\beta^2 (\alpha - \bar{x})^3} \alpha^2_2 \, .
\end{aligned}
\end{equation}

We want to bring the NVE, \cref{eq: NVE_x}, in a more convenient form for the application of the Kovacic algorithm \footnote{Recall that we do not know the exact form of $\bar{x}$ explicitly.}. With
that in mind, we consider a new variable introduced via 
\begin{equation}
\bar{x} = z 
\end{equation}
and under this change, the NVE \cref{eq: NVE_x} now becomes 
\begin{equation} \label{eq: NVE_x_Kovacic}
\begin{aligned}
\dot{z}^2 \frac{d^2 \mathcal{X}}{dz^2} &+ (\ddot{z} + \dot{z}^2 \mathcal{P}) \frac{d \mathcal{X}}{dz} \\ 
&+ \left( \mathcal{Q}_1 \ddot{z} + \mathcal{Q}_2 \dot{z}^2 + \mathcal{Q}_3 \right) \mathcal{X} = 0 \, ,
\end{aligned}
\end{equation}
with the $\mathcal{P}, \mathcal{Q}_1, \mathcal{Q}_2, \mathcal{Q}_3$ being evaluated on $\bar{x}=z$.

We can use the worldsheet equations of motion \cref{eq: virasoro_spec_1} on the straight line solution and on $x=\bar{x}=z$ to solve for $\dot{z}^2$. This yields
\begin{equation}
\begin{aligned}
\dot{z}^2 = &4 \frac{-\mu + (z-\alpha)(z-\beta)z}{\alpha-z} \\
				&\left(E^2 - J^2 + \frac{z^2 - z \beta+\frac{\mu}{z-\alpha}}{\beta^2} \right) \alpha^2_2\, ,
\end{aligned}
\end{equation}
and use the equations of motion for $x$, \cref{eq:xeom_1}, evaluated again on the straight line solution and on $x=\bar{x}=z$ to re-express $\ddot{z}$. We get 
\begin{equation}
\begin{aligned}
&\ddot{z} = \frac{2}{(z-\alpha )^2 (\alpha -z)}(z (z-\alpha ) (z-\beta )-\mu) \left(\vphantom{\frac{1}{2}} \right. \\
&\left.+ \frac{\mu +\beta  (z-\alpha )^2}{\beta ^2} \alpha^2_2 - \frac{4(\alpha  \beta +3 z^2-2 z (\alpha +\beta ))}{\beta ^4 (z-\alpha )}  \right. \\
&\left. \left((z-\alpha ) \left(\beta ^2 (E^2-J^2)+z(z-\beta) \alpha^2_2  \right)-\alpha^2_2 \mu \right){}^2  \right. \\
&\left.+4 \left( E^2 -J^2 + \frac{z^2 - \beta z + \frac{\mu}{-z+\alpha}}{\beta^2} \right)^2  \right. \\
&\left.(z (z-\alpha ) (z-\beta )-\mu) 
\vphantom{\frac{1}{2}} \right)
\end{aligned}
\end{equation}

We follow the analytic Kovacic algorithm, which has been very thoroughly reviewed in \cite{Filippas:2019ihy}, and we deduce that no combination of the parameters $\{a,b,c \}$ provides a Liouville integrable solution of the NVE which suggests that the system is non-integrable for general values that characterize the $L^{a,b,c}$ model. 
		\subsubsection{\textbf{S\lowercase{olving the Lagrangian equations of motion}}}
While the non-integrability of a system does not imply chaos necessarily, chaotic dynamics is indicative of the absence of integrability. In this and the next section we will perform numerical analysis of the equations of motion for the extended string we have considered and allow the system to evolve in time. This time evolution reveals chaotic dynamics.

The equations of motion for $x(\tau)$ and $\theta(\tau)$ are coupled in general as we saw, and though we are not able to find exact analytic solutions we can solve them numerically. 

We choose to study the $L^{1,7,5}$ manifold. We can immediately see that we get  $d=3$ using \cref{eq: defd}. We have the freedom to set any of the $\alpha, \beta, \mu$ constants to any non-zero value and we choose to set $\mu=1$. We solve the system of equations described by the first equation in \cref{eq: manifolddef} as well as the relations described in \cref{eq: usefuleqn} to determine the values of $\alpha, \beta, x_1, x_2, x_3$. We obtain $\alpha=2.854, \beta=3.188, x_1=0.1192, x_2=2.347, x_3=3.575$. We can also determine the period of the coordinate $\zeta$, which is equal to $\breve{\zeta}=0.9782$. We solve both of the equations of motion, \cref{eq:thetaeom_1,eq:xeom_1}, numerically by choosing as initial conditions $\theta = 0.8$ and $x=0.3$. For the winding of the string along the two $U(1)$ angles inside the $L^{1,7,5}$ manifold we choose $\alpha_1=1$ and $\alpha_2=2$. The initial choice for $x(\tau)$ is such that it lies between the two smallest roots of the cubic equation $\Delta_x=0$ as required. We let the system evolve in time and we plot $\sin \theta$ in a similar manner to the $AdS_5 \times T^{1,1}$ case \cite{Basu:2011di}. The result is presented in \Cref{fig: eleomsltn}. 

An interesting special case of the $L^{a,b,c}$ models is to consider $b=c$ \cite{Benvenuti:2005ja,Franco:2005sm} with the usual condition $a \leq b$. This special class of models has been dubbed generalized conifolds. The $L^{a,b,b}$ case is equivalent to the $L^{a,b,a}$ under some trivial reorderings as is explained in \cite{Franco:2005sm}. 

We study the generalized conifold given by $L^{1,5,5}$. Following the same steps as before for the $L^{1,7,5}$ we obtain that $d=1$ and we set again $\mu=1$. The values that characterize the model for the constants $\alpha, \beta$ are $1.931, 3.315$ respectively. The three roots of the cubic $\Delta_x=0$ are given by $x_1=0.81825, x_2=1.566, x_3 = 3.498$. The coordinate $\zeta$ ranges from $0$ to $\breve{\zeta}=1.501$ and the remaining needed values for the numerical solution of the equations of motion are the same as in the $L^{1,7,5}$ example. As we did previously, we show the time evolution of the string motion in \Cref{fig: eleomsltn}.

In both cases, the string motion exhibits chaos. 
		\subsubsection{\textbf{T\lowercase{he} L\lowercase{yapunov exponent}}}
 A characteristic feature of chaos is the sensitivity of a system to a specific choice for initial conditions. Having said that, we discuss the largest Lyapunov exponent (LLE). The sensitivity on the initial conditions can be phrased in the following way: we can consider any point  in the phase space of the theory which we call $X$. There exists at least one point which lies in an infinitesimally close distance to that point and that diverges from it. The said distance is denoted by $\Delta X(X_0, \tau)$ and is a function of the initial position. The largest Lyapunov exponent is a characteristic quantity that quantifies the rate of separation of such closely laying trajectories in the theory's phase space. It is given by 
\begin{equation}
\lambda = \lim_{\tau \rightarrow \infty} \lim_{\Delta X \rightarrow 0} \left(\frac{1}{\tau} \log \frac{\Delta X(X_0,\tau)}{\Delta X(X_0, 0)} \right) \, .
\end{equation}

\begin{figure}[t]
\centering
\includegraphics[scale=0.65]{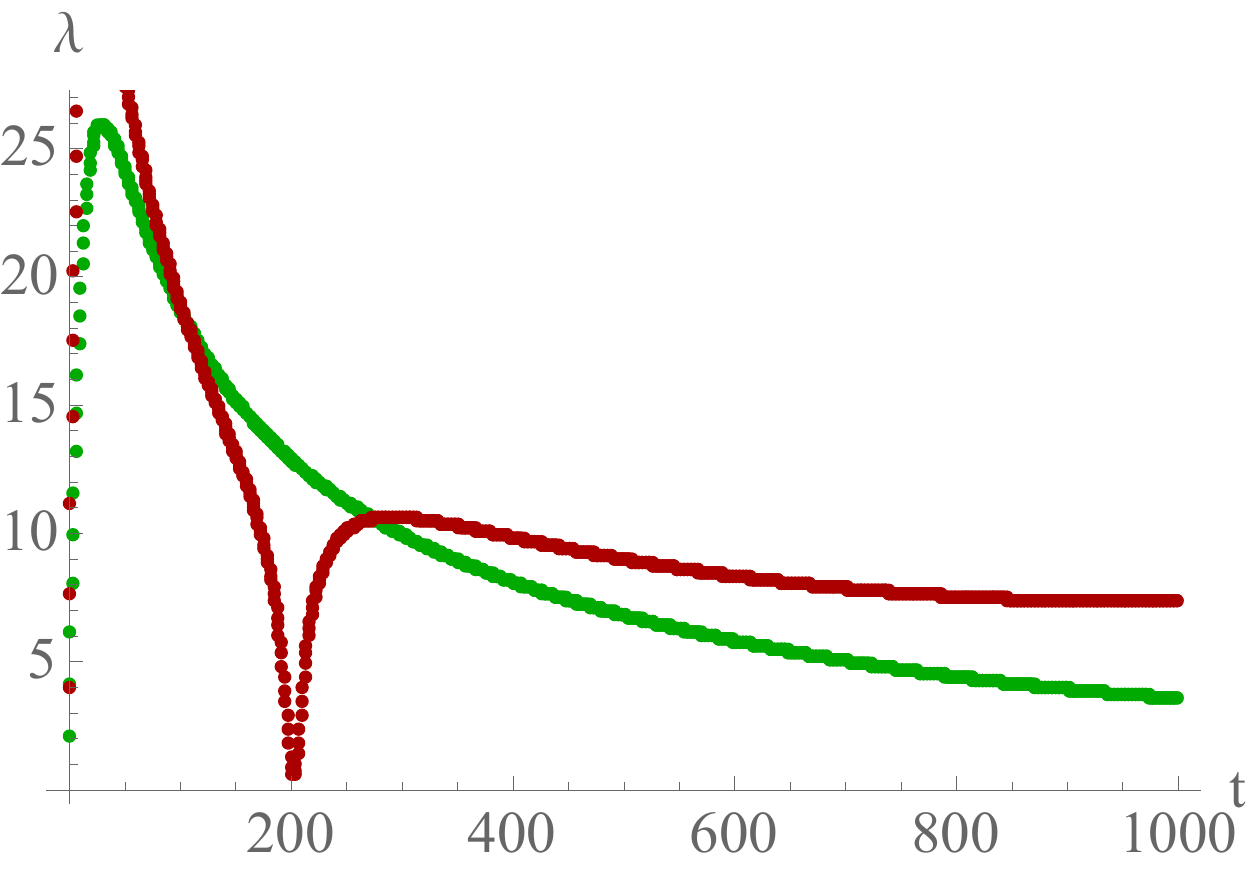} 
\centering
\caption{\label{fig: lyapunov} The Lyapunov index for the extended string motion in the $L^{1,7,5}$ manifold (red) with $x(0)=0.12, \theta(0)=\pi/4$ and the $L^{1,5,5}$ model (green) with $x(0)=0.2, \theta(0)=0.1$. For the former, we observe a convergence with $\lambda \approx 7.3$ and for the latter $\lambda \approx 3.5$.}
\end{figure}

We compute the LLE for the systems under consideration. We expect that as we dynamically evolve the system in time and for a chaotic motion, $\lambda$ will converge to some non-zero positive value and fluctuate around that particular value. We have verified that such is the case for the extended string given by \cref{eq: string_soliton} that is moving in the $L^{a,b,c}$ manifolds and the result of the computation is shown in \Cref{fig: lyapunov}. 
\section{BPS \lowercase{mesons and point-like strings}} \label{sec: geodesics}
 In \cite{Benvenuti:2005ja} the authors identified the angle conjugate to the R-symmetry and argued that the BPS geodesics resulting from these point-like string modes are compared to the BPS mesons of the quiver theory. Below we study the (non)-integrability of point-like strings.
	\subsection{Point-like string motion}
We have examined the dynamics of extended string configurations so far. Now we turn our attention to the point-like limit of the string. This limit is obtained very straightforwardly. The change, compared to the previous case, is that the string now is not wrapping the two coordinates inside the $L^{a,b,c}$; simply put we set $\alpha_1=\alpha_2=0$ in \cref{eq: string_soliton}. 

The Lagrangian can be obtained readily by the previous expression. It is given by:
\begin{equation} \label{eq: string_lagrangian_point}
\begin{aligned}
&\mathcal{L} = \dot{t}^2 - \dot{\zeta}^2 - \frac{\rho^2}{4} \left(  \frac{\dot{x}^2}{\Delta_x} + 4 \frac{\dot{\theta}^2}{\Delta_{\theta}} \right) 
\end{aligned}
\end{equation}

The equations of motion that follow from the Lagrangian are: 
\begin{subequations}
\begin{equation}
\begin{aligned}
&\frac{2(\Delta_{\theta} -x ) \ddot{x}}{\Delta_x} = - \frac{\dot{x}^2}{\Delta_x} - \frac{4 \dot{\theta}^2}{\Delta_{\theta}} \\
&+ \frac{(\Delta_{\theta}-x)(\alpha \beta - 2(\alpha+\beta)x + 3 x^2)\dot{x}^2}{\Delta^2_x} \\
&+2 \frac{\dot{x} + (\alpha-\beta) \sin(2 \theta) \dot{\theta}}{\Delta_x} \dot{x} 
\end{aligned}
\end{equation}
\begin{equation}
\begin{aligned}
&\frac{2(\Delta_{\theta} -x ) \ddot{\theta}}{\Delta_{\theta}} = \frac{2 \dot{x} \dot{\theta}}{\Delta_{\theta}} - (\alpha - \beta) \sin(2 \theta) \left( \frac{\dot{\theta}^2}{\Delta_{\theta}} \right. \\
&\left.+ \frac{\dot{x}^2}{4 \Delta_x} - \frac{2 \dot{\theta}^2}{\Delta_{\theta}} + \frac{1}{2 \Delta^2_{\theta}} ( \alpha + \beta  \right. \\
&\left.+ (\alpha-\beta) \cos(2 \theta)-2x) \dot{\theta}^2
\right)
\end{aligned}
\end{equation}
\end{subequations}

The Virasoro conditions that constrain the equations of motion for the point-like string read
\begin{subequations} \label{eq: virasoro_spec_point}
\begin{equation} \label{eq: virasoro_spec_1_point}
\begin{aligned}
2T_{\tau \tau} = 2T_{\sigma \sigma} = - \dot{t}^2 &+ \dot{\zeta}^2 \\ 
&+ \frac{\rho^2}{4} \left(  \frac{\dot{x}^2}{\Delta_x} + 4 \frac{\dot{\theta}^2}{\Delta_{\theta}} \right) = 0,
\end{aligned}
\end{equation}
\begin{equation} \label{eq: virasoro_spec_2_point}
\begin{aligned}
T_{\tau \sigma}=T_{\sigma \tau} = 0.
\end{aligned}
\end{equation}
\end{subequations}

We can, of course, express the system in a Hamiltonian formalism. The canonical conjugate momenta are given by 
\begin{equation}
\begin{aligned}
p_t &= 2 \dot{t}, \quad 									&p_{\zeta} &= -2 \dot{\zeta}, \\
p_x&=-\frac{\rho^2}{2 \Delta_x} \dot{x}, \quad 	&p_{\theta} &= - \frac{2 \rho^2}{\Delta_{\theta}} \dot{\theta},
\end{aligned}
\end{equation}
and the associated Hamiltonian density is equal to
\begin{equation}
\begin{aligned}
&\mathcal{H} = \frac{1}{4 \rho^2} \left[ \rho^2 (p^2_t - p^2_{\zeta}) - 4 \Delta_x p^2_x - \Delta_{\theta} p^2_{\theta} \vphantom{\frac{1}{2}} \right]  \, .
\end{aligned}
\end{equation}

The invariant plane of solutions on which the equations of motion are satisfied is given by:
\begin{equation}
\{ x = x_0, \dot{x} = \ddot{x} = 0 , \theta = \theta_0, \dot{\theta} = \ddot{\theta} = 0  \} \, ,
\end{equation}
alongside with the simple solutions 
\begin{equation}
t = E ~ \tau + c_1, \quad \zeta = J ~ \tau + c_2 \, .
\end{equation}

It is quite straightforward to see that if we expand $t = E ~ \tau + c_1 + \varepsilon \tilde{t}$ as well as $\zeta = J ~ \tau + c_1 + \varepsilon \tilde{\zeta}$, with $\varepsilon \rightarrow 0$, we are led to the NVEs for the $t$ and $\zeta$ respectively. Both of them admit Liouville integrable solutions. 

We can also fluctuate the $x$-coordinate on the invariant plane as $x = x_0 + \varepsilon \mathcal{X}$ with $\varepsilon \rightarrow 0$ to obtain 
\begin{equation}
\frac{(\alpha -\beta ) \cos (2 \theta_0 )+\alpha +\beta -2 x_0}{x_0 (x_0-\alpha ) (x_0-\beta )-\mu } \ddot{\mathcal{X}} = 0
\end{equation}
which also has Liouville integrable solutions. 

Similarly, we can obtain the NVE for the $\theta$-coordinate. We expand as $\theta = \theta_0 + \varepsilon \vartheta$ in the limit $\varepsilon \rightarrow 0$ and derive
\begin{equation}
\frac{2 \left(\alpha  \cos^2(\theta_0)+\beta  \sin^2(\theta_0)-x\right)}{\alpha  \cos^2(\theta_0)+\beta  \sin^2(\theta_0)} \ddot{\vartheta} = 0
\end{equation}
which has Liouville integrable solutions as in the previous cases.
	\subsection{C\lowercase{hanging coordinates and the} R-\lowercase{symmetry angle}} \label{sec: new_coord}
Let us briefly describe the change of variables that was introduced in \cite{Benvenuti:2005ja}. It is given by $y=\cos(2\theta)$. Moreover, the said change of variables makes the comparison between BPS geodesics and mesons straightfroward. In order to be able to make a statement for the operators of the boundary quiver, one needs to know the angle conjugate to the R-symmetry. This was also obtained in the aforementioned paper and it reads: 
\begin{equation}
\Omega_{R} = 3 \zeta + \phi + \psi\, ,
\end{equation}
Now one is able to re-express the geometry \cref{eq: Labc_part1,eq: Labc_part2} in terms of this angle and the coordinate $y$. Here we choose not to do that, however we find it useful and illuminating to have this expression explicitly in order to be able to draw conclusions directly using our coordinate system - $\phi, \psi$.
	\subsection{BPS \lowercase{mesons from strings}}
It has been shown that BPS mesons correspond to the BPS geodesics \cite{Benvenuti:2005ja}. These geodesics are such that $x=x_0$ and $y=y_0$. This can be easily translated into the following statement in our coordinates $x=x_0$ and $\theta=\theta_0$, where the constants are such that they respect the ranges we have discussed. Moreover, it was argued that the necessary minimization of the Hamiltonian is achieved for $\dot{\phi} = \dot{\psi} = 0$. This is the same string configuration that we examined above by taking the point-like limit of the string.
\section{E\lowercase{pilogue}}
In this work we considered the motion of an extended string that is localized at the centre of the $AdS_5$ space and is wrapping two $U(1)$ angles inside the $L^{a,b,c}$ space. We showed that the dynamics of that particular string configuration is non-integrable, since the Kovacic algorithm fails to provide a solution to the fluctuation equations. We also studied the coupled equations of motion that were derived from the Lagrangian and solved them numerically. The time evolution indicates chaotic dynamics for the string which is another characteristic signature of non-integrability. Having observed the chaotic dynamics, we computed the largest Lyapunov exponent which was found to converge to some positive value. 

Since type IIB string theory in the AdS$_5 \times L^{a,b,c}$ vacuum is holographically dual to the $\mathcal{N}=1$ quiver gauge theories and we managed to argue that the string picture in the bulk has a translation to the field theory operators, we have, essentially, argued that these particular quiver gauge theories are non-integrable on general grounds. 

We also examined the dynamics of a string configuration in the point-like limit and we managed to derive Liouville integrable solutions to the NVE. This is another situation where the integrability of the extended string motion appears to be a much more stringent statement than the integrability of moving particles.

Finally, our work here combined with the results obtained previously in \cite{Basu:2011di,Basu:2011fw} suggest that the classical string motion in the $AdS_5 \times \mathcal{M}^5$ vacuum, with $\mathcal{M}^5$ being a five-dimensional Sasaki-Einstein manifold, is non-integrable. 
\section*{A\lowercase{cknowledgements}}
I am grateful to N. J. Evans, D. Giataganas and T. Nakas for helpful discussions. I also thank K.Filippas for an elucidating discussion on the Kovacic alogrithm as well as D. Giataganas and C. Nunez for a careful reading and enlightening comments on the final draft of this work. 
\bibliographystyle{apsrev4-1} 
\bibliography{lit.bib}

\end{document}